\documentclass{elsart}

\usepackage{natbib}

 \usepackage{graphics}
\usepackage{graphicx}

 \usepackage{amssymb}

\newcommand{\BibTeX}{ \textrm{B\kern-.05em\textsc{i\kern-.025em b}\kern-.08em
    T\kern-.1667em\lower.7ex\hbox{E}\kern-.125emX} }

\begin{document}

\begin{frontmatter}



\title{Searching for Main-Belt Comets Using the Canada-France-Hawaii Telescope Legacy Survey}


\author[label1,label2]{Alyssa M. Gilbert} and
\author[label1]{Paul A. Wiegert}

\address[label1]{Department of Physics \& Astronomy, The University of Western Ontario,
				London, ON N6A 3K7 (Canada)}
\address[label2]{Corresponding Author E-mail address: alyssa.gilbert@uwo.ca}

\begin{center}
\scriptsize
Copyright \copyright 2008 Alyssa M. Gilbert and Paul A. Wiegert
\end{center}


\begin{abstract}

The Canada-France-Hawaii Telescope Legacy Survey, specifically the Very Wide segment of data, is used to search for possible main-belt comets. In the first data set, 952 separate objects with asteroidal orbits within the main-belt are examined using a three-level technique. First, the full-width-half-maximum of each object is compared to stars of similar magnitude, to look for evidence of a coma. Second, the brightness profiles of each object are compared with three stars of the same magnitude, which are nearby on the image to ensure any extended profile is not due to imaging variations. Finally, the star profiles are subtracted from the asteroid profile and the residuals  are compared with the background using an unpaired T-test. No objects in this survey show evidence of cometary activity. The second survey includes 11438 objects in the main-belt, which are examined visually. One object, an unknown comet, is found to show cometary activity. Its motion is consistent with being a main-belt asteroid, but the observed arc is too short for a definitive orbit calculation. No other body in this survey shows evidence of cometary activity. Upper limits of the number of weakly and strongly active main-belt comets are derived to be $630\pm77$ and $87\pm28$, respectively. These limits are consistent with those expected from asteroid collisions. In addition, data extracted from the Canada-France-Hawaii Telescope image archive of main-belt comet 176P/LINEAR is presented.

\end{abstract}
 
\begin{keyword}
Asteroids \sep Comets \sep Comet 176P
 
 
\end{keyword}


\end{frontmatter}


\section{Introduction}
Historically, asteroids and comets have been thought as two separate populations in the Solar System. Having differing volatile fractions, a common distinction between the two is comets display a coma, while asteroids do not. In addition, the two populations differ dynamically. A standard criterion used to differentiate between a cometary or asteroidal orbit is the Tisserand parameter with respect to Jupiter,

\begin{equation}
T_J=\frac{a_J}{a}+2\cos i\sqrt{\frac{a}{a_J}(1-e^2)},
\end{equation}

where $a_J$ is the semi-major axis of Jupiter, and $a$, $e$, and $i$ are the semi-major axis, eccentricity, and inclination of the object, respectively. Comets typically have $T_J<3$, while asteroids have $T_J>3$.

Recently, surveys such as those discussed in \citet{FJS2005}, \citet{J2005}, and \citet{L2008} have discovered a significant number of asteroids in comet-like orbits. There are also objects in asteroid-like orbits that show bursts of cometary activity ($7968=133$P/Elst-Pizzaro; \cite{E1996}) or are associated with meteor streams (near Earth object 3200 Phaethon; \cite{W1983}). These observations have made the boundary between comets and asteroids less obvious. Intermediate objects may be comets that are extinct, dormant, or dead. Conversely, they may be asteroids with higher volatile content.

Objects that display activity but are in asteroid-like orbits are known as main-belt comets (MBCs) or activated asteroids (AAs) \citep{HJ2006}. These objects are most likely native to the asteroid belt \citep{F2002} and may be activated by a collision with a small body \citep{B1998,T2000}. \cite{HJ2005,HJ2006b} carry out the first survey to search for MBCs and, as of this writing, only three MBCs have been found: 133P/Elst-Pizzaro \citep{E1996}, P/2005 U1 \citep{R2005} and 176P/LINEAR \citep{HJ2006}.

In this paper, we present a study using the Canada-France-Hawaii Telescope (CFHT) Legacy Survey data to search for cometary activity in dynamically asteroidal bodies. The first (smaller) data set is subjected to stringent tests designed to detect weak activity. These objects are examined by comparing their full-width-half-maximum (FWHM) measurements and brightness profiles to those of stars of similar magnitude. If the profile of the asteroid is broader than those of the stars, it may indicate the presence of a coma. For such objects, the star profiles are subtracted from the profile of the asteroid, and an unpaired T-test is used to compare the residuals with the background to determine whether they are significantly different. A second (larger) data set is visually examined for evidence of stronger cometary activity. Using results from both methods, upper limits are derived for the number of weakly and strongly active MBCs and are compared to those expected from collisional activation of asteroids. Finally, we present measurements of the known MBC 176P/LINEAR in the CFHT archive.

\section{Observations and Data Reduction}
All images were acquired with MegaCam on the 3.6-m CFHT in Hawaii. The images were taken as part of the Very Wide (VW) segment of the CFHT Legacy Survey (CFHTLS; \cite{J2006,K2009}). The asteroids forming the first data set were observed on 2004 December $15-16$, 2005 January $16-17$, and 2006 May $1-2$ and $25-26$. All observations were taken in either the g' or r' filters, with exposure times of 90~s and 110~s, respectively. The limiting magnitude for a 90\% probability of a $3\sigma$ detection in the g' filter was 22.5, and 21.75 for the r' filter. The average seeing size was $1^{\prime\prime}$ in both filters.

The observations for the second data set were taken on various dates between August 2003 and January 2008. The exposure times for the g' and r' filters were $70-110$~s and $110-180$~s, respectively, yielding comparable limiting magnitudes to the observations above.

The images were pre-processed using the Elixer pipeline \citep{MC2004}. A fine astrometric correction was applied by the TERAPIX data-processing center, and the data was stored at the Canadian Astronomical Data Centre (CADC). Source Extractor \citep{BA1996} and additional software designed for detecting moving bodies were used to find asteroidal objects. A more detailed explanation of the observations and data reduction process for the smaller data set are found in \citet{W2007}, which used that data to investigate the size distribution of kilometer-sized main-belt asteroids (MBAs) as a function of color.

For both data sets, each field was typically observed three times on the first night, approximately 45 minutes apart, and once the following night. This allowed for a reasonable determination of orbit parameters for main-belt objects. The observations were taken at opposition and were not selected based on their possible main-belt content. Unlike other searches for MBCs, which focus on individual objects \citep{C1996,L1990} or asteroid families \citep{HJ2006}, this was a relatively unbiased survey of the main-belt. This allowed for a determination of the upper limit of MBCs expected for the whole main-belt, rather than for a sub-population. In this study, observations from the first night were utilized to search for cometary activity. When a second night of data was acquired, it was used for orbit refinement.

\section{Data Analysis}
\subsection{Three-Level Analysis}
To determine whether an asteroid showed evidence of cometary activity, three levels of analysis were chosen, each refining the number of possible MBC candidates. The investigation began with 1468 asteroids that were determined to be small bodies in the main-belt by visual inspection of image triplets \citep{W2007} (each of these objects were also included in the visual investigation discussed below). The analysis was limited to objects with magnitudes less than 21.5 since the wings of the seeing profiles of fainter asteroids became too noisy for reliable examination in subsequent stages. Applying this limit left 952 objects to be examined. 

In the first level of analysis the FWHM measurements of the objects and stars were compared. Each asteroid image was rotated by the angle of the direction of motion, calculated from the RA and Dec coordinates, such that the direction of motion lied on the horizontal axis. The FWHM was measured perpendicular to the direction of motion using IRAF's IMEXAMINE tool. The FWHM along the direction of motion of the asteroid was not measured, since the images were trailed.

All stars in the image were divided into magnitude bins (0.25-mag wide) and the median FWHM of each bin was calculated. The FWHMs of each asteroid and stars in the same magnitude bin were compared. If the FWHM of the asteroid was greater than 110\% of the FWHM of the stars in two or more observations the asteroid was passed to the next analysis level. The arbitrary limit of 110\% was chosen to provide a substantial number of asteroids with the highest FWHMs. Of the initial 952 asteroids, 415 passed this test.

In the second level of analysis, the brightness profile of each asteroid was compared to three stars of similar magnitude ($\pm$0.25~mag) on the same image and CCD chip. This eliminated the effect of imaging variances across the CCD. The brightness profiles were created from a $20\times20$ matrix of pixel counts. The counts of the columns were averaged, the background was subtracted and the columns were plotted as a function of the line number. Each profile was subsequently scaled to a height of unity. 

The profiles of the asteroids were examined for interesting features (i.e., extended wings) when compared visually to the stars. Although this process was subjective, the intention was to select objects having scaled counts in the wings more than 0.05 higher than the stars. An example of a typical extended profile that passed this test is shown in Fig.~1. There were 65 asteroids that showed a similar feature to Fig.~1, and subsequently passed this test.

[Figure 1]

In the third level of analysis, the three star profiles were subtracted from the asteroid profile and an unpaired T-test was used to compare the residuals with the background noise. Using this method, 14 objects had residuals that were significantly different (i.e., $T>2$ for $\alpha=0.05$) from the background in at least two observations.

The images of those 14 objects were examined more carefully. All of the asteroids were either overlapping or near a background object,  internal reflections, or artifacts due to bright stars, thus making the object appear more extended. Therefore, no asteroids passed the final analysis level, or could be considered good MBC candidates.

\subsection{Visual Investigation}
The larger data set consisted of a total of 11615 objects which were visually inspected (as in \cite{HJ2006}). Some of these objects were outside the main-belt and were excluded from this analysis, leaving 11438 objects to be examined. A single known comet was found among the non-main-belt detections: active Centaur 166P/NEAT ($a=13.88$~AU, $q=8.56$~AU, $e=0.383$, $i=15.36^o$) in a distinctly active phase.

An unknown object, observed on 2007 September 14, was discovered to show cometary activity. Its on-sky motion was consistent with being an MBC, but the orbit was poorly constrained. No known comets were found in the same region using MPChecker and Jet Propulsion Laboratory (JPL) Horizons. A report to the Minor Planet Center (MPC) did not produce any response linking it to another known body, either cometary or asteroidal; therefore, these are concluded to be the discovery observations of this object, pending further information from the MPC. It remains unclear whether this object is an MBC.

Three observations of this object were taken over a 79-minute arc on one night and a fourth was taken approximately 48 hours later. The apparent magnitude varied from 21.2 to 21.6 in the r' filter. This yielded an absolute magnitude of 16.2 or a diameter, $D$, of $1.5-3.5$~km for albedos ranging from 0.05 to 0.25 (assuming the main-belt orbit calculated below). A faint, but distinct, coma/tail of approximately 10$^{\prime\prime}$ in length was visible in all images (Fig.~2).

[Figure 2]

The images were retrieved from the CFHTLS archive at the CADC and searched for moving objects. Since this process was completed 12 months following the observations, telescopic follow-up at the time of discovery was impractical. A search of the CFHT archive found no other images that were expected to contain the comet.

The observations were taken near opposition (phase angle of $4^o$). The motion over 48 hours was consistent with an MBA and, assuming the usual Vaisala circular orbit, parameters of $a\sim3.9$~AU and $i\sim13^o$ were derived. However, the arc was too short to conclude this definitively, and the motion could also be fit with slightly smaller residuals (RMS 0.04$^{\prime\prime}$ versus 0.14$^{\prime\prime}$) by a more traditional cometary orbit with $a\sim6$ AU, $e\sim0.6$ and $q\sim2.5$~AU. The nominal best fit cometary orbit would have passed perihelion in November 2006, about a year before the object was observed in September 2007.

Statistically, we expect this object is more likely to be a true comet rather than an MBC because of the relative rarity of the latter \citep{HJ2006}. Its true nature will only be revealed once it is recovered. The observations, as reported to the MPC, appear in Table~1.

[Table 1]

The unknown comet and 166P were used to verify the three-level analysis technique. Figures~3 and 4 show the brightness profiles of 166P and the unknown comet compared to three stars, respectively. Both objects passed the three levels of analysis. No other objects in this data set show cometary activity.

[Figure 3]

[Figure 4]

\section{Results}
\subsection{Expected Number of Active MBCs}
Two of the three known MBCs (133P/Elst-Pizzaro and 176P/LINEAR) are members of the Themis family of asteroids in the outer asteroid belt ($3.05<a<3.22$~AU, $0.12<e<0.19$, $0.69^o<i<2.23^o$; \cite{Z1990}), while P/2005 U1 is just outside this family because of its slightly higher eccentricity ($e=0.25$). The unknown comet from this survey resides in the outer belt ($a\sim3.9$~AU), though its inclination of $13^o$ is incompatible with the Themis family.

Inside 3.2~AU, interior temperatures of asteroids are higher and may prevent the survival of ice \citep{HJ2006}. Therefore, when deriving the expected numbers of MBCs, we initially assume that MBCs must be located in the outer belt, beyond 3.0~AU.

The objects in this study range in size down to $D\sim1$~km and $D\sim1.5$~km for the large and small data sets, respectively. From the cumulative size distribution given in \citet{C2004}, the number of asteroids with $D\gtrsim1$~km is $\sim1\times10^6$, while for $D\gtrsim1.5$~km is $\sim6\times10^5$. Using the de-biased number of asteroids as a function of magnitude derived by \citet{JM1998}, $\sim30\%$ of asteroids with absolute magnitude less than 16 have $3.0<a<5.0$~AU. Extrapolating to smaller sizes, we estimate $\sim30\%$ of all asteroids ($\sim3\times10^5$ with $D\gtrsim1$~km and $\sim1.8\times10^5$ with $D\gtrsim1.5$~km) are in the outer belt.

In the first data set, 216 of 952 objects fall within $3.0<a<5.0$~AU, which is consistent with the 30\% figure stated above. Assuming MBCs only occur in the outer belt gives an upper limit of 1/216 (0.46\%). Using the number of asteroids expected in this region ($1.8\times10^5$) gives an upper limit of $830\pm86$ currently weakly active MBCs, where the uncertainty is $\pm3\sigma$ (assuming a Poisson distribution).

Using the 11438 objects from the second data set, the limit is constrained further. Although visual inspection may be insufficient to detect faint comae associated with weak activity, it is capable of determining stronger activity, as demonstrated by the detection of two moderately active comets. Of the 11438 objects, 2667 have orbits with $3.0<a<5.0$~AU, giving an upper limit of $0.037\%$, or $110\pm31$ strongly active MBCs. This limit agrees with the the $15-150$ range derived by \citet{HJ2006}.

If MBCs can occur anywhere in the main-belt, the active fractions for weakly and strongly active MBCs change to 0.11\% and 0.009\%, respectively. Using the total number of $6\times10^5$ for asteroids in the whole main-belt with $D\gtrsim1.5$~km, the upper limit for weakly active MBCs becomes $630\pm77$. For strongly active MBCs, the upper limit is $87\pm28$, assuming there are $1\times10^6$ asteroids with $D\gtrsim1$~km.

\subsection{Comparison with Collisional Lifetimes}
The upper limits derived for MBCs are compared with those expected from collisional activation of MBAs. The activation timescale for MBCs is unknown, but if collisional, is most likely comparable to the time between significant sub-catastrophic impacts. This rate is approximately the destructional lifetime, which is derived by \citet{C2004} to be $10^8$~years for asteroids with $D\sim1-5$~km (the size range of the majority of our objects). This is chosen as a first estimate of the collisional activation interval of MBCs.

Assuming there are $3\times10^5$ asteroids in the outer main-belt with $D\sim1-5$~km, and the active lifetimes of these objects after a collision is $10^{3-4}$~years \citep{HJ2006,DG2008}, there should be $3-30$  active asteroids due to collisions. This is consistent with the upper limit derived for active MBCs. This derivation assumes MBCs remain consistently active for thousands of years after activation, though observations indicate their activity is intermittent (\cite{H2007}; see below).

\subsection{Main-Belt Comet 176P/LINEAR}
The CFHT image archive was searched for observations of the three known MBCs. Only one was found, 176P/LINEAR, which was observed on 2005 June 10 and 2007 January 15. For analysis, three 60-s observations from 2005 June 10 were used, with seven and two minutes between observations, respectively. There were three 150-s observations from 2007 January 15, taken $\sim40$~minutes apart.

The profiles of 176P and stars of similar magnitude were compared, and the unpaired T-test  was performed on both sets of observations. None of the images passed the T-test, indicating 176P was not active during these times.

\citet{H2007} determined 176P was active during November and December 2005, just after perihelion. However, it was observed to be inactive shortly before and after this, in October 2005 and February 2006. With an orbital period of 5.71~years, 176P would be expected to be inactive during June 2005, when it was approaching perihelion, and January 2008, when it was about half way between aphelion and perihelion. Figure~5 shows the orbital position of the active and inactive phases of 176P.

[Figure 5]

\section{Conclusions}
In this study, 952 asteroids are examined using three levels of analysis and 11438 are inspected visually. In the larger set, one unknown object  is discovered to display cometary activity. Since the orbit determination for this object has large uncertainties, it is possible it is an MBC. Statistically, however, it is more likely to be a true comet. The unknown comet, as well as 166P, are used to successfully test the three-level analysis technique. No other object shows evidence of cometary activity, so there are zero or one MBC candidates in the survey.

An upper limit is derived for weakly active MBCs to be $830\pm86$ for objects with $D\gtrsim1.5$~km and $3.0<a<5.0$~AU. The larger sample of visually inspected objects in the outer main-belt with $D\gtrsim1$~km produces an upper limit of $110\pm31$ strongly active MBCs. Expanding to the whole main-belt, upper limits are $630\pm77$ and $87\pm28$ for weakly and strongly active MBCs, respectively. These agree with previous limits derived by \citet{HJ2006} and are in good agreement with what is expected from collisional activation of asteroids. It is concluded that the activity of MBCs is consistent with activation caused by collisions with other small main-belt objects.

\ack
We would like to thank David L. Clark for his help searching the CFHT archives for images of the three known MBCs and the unknown comet. This work is supported in part by the National Science and Engineering Council of Canada. This work is based on observations obtained with MegaCam, a joint project of CFHT and CEA/DAPNIA, at the CFHT which is operated by the National Research Council (NRC) of Canada, the Institut National des Science de l'Univers of the Centre National de la Recherche Scientifique (CNRS) of France, and the University of Hawaii. This work is based in part on data products produced at TERAPIX and the CADC as part of the CFHTLS, a collaborative project of NRC and CNRS.

\label{lastpage}


\bibliography{bibliography.bib}
\bibliographystyle{elsart-harv}

\vfill

\newpage

\begin{table}
\begin{center}
 \textbf{Observations of the Unknown Comet}
\begin{tabular}{llll}
\hline 
\hline
Obs. Date (UT) & RA & Dec & mag \\ \hline
2007 09 14.27098 & 22 51 27.56 & +09 59 35.9 & 21.3 \\
2007 09 14.30181 & 22 51 26.38 & +09 59 29.6 & 21.2 \\
2007 09 14.32586 & 22 51 25.45 & +09 59 24.8 & 21.2 \\
2007 09 16.29386 & 22 50 12.88 & +09 52 33.4 & 21.6 \\ \hline
\end{tabular}
\caption[Observations of the Unknown Comet]
	{
	\label{files}	
	\label{lasttable}		
	Four observations of the unknown comet discovered in the CFHTLS data. Each image was taken in the r' filter and had exposure times of 180~s.
	}
\end{center}
\end{table}

\clearpage

\begin{figure}[t!]
\begin{center}
\includegraphics[width=6in]{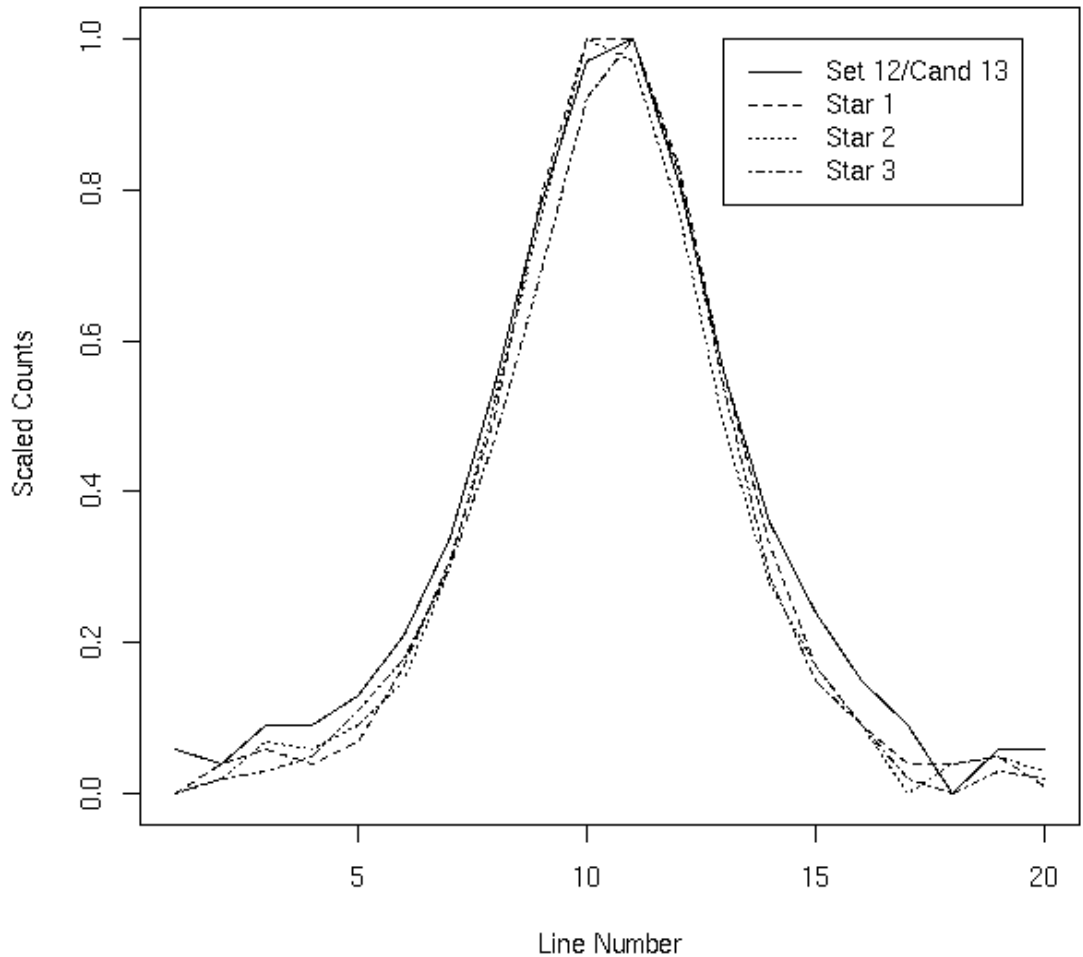}
\caption[Example Object and Star Profiles]{
	\label{fig1}
	\label{lastfig}			
	The solid black line represents the brightness profile of an asteroid in the sample, while the three dotted lines represent the profiles of three comparison stars on the same chip. The asteroid shows slightly more counts in the wings, possibly due to a coma.
	}
\end{center}
\end{figure}

\begin{figure}[t!]
\begin{center}
\includegraphics[width=6in]{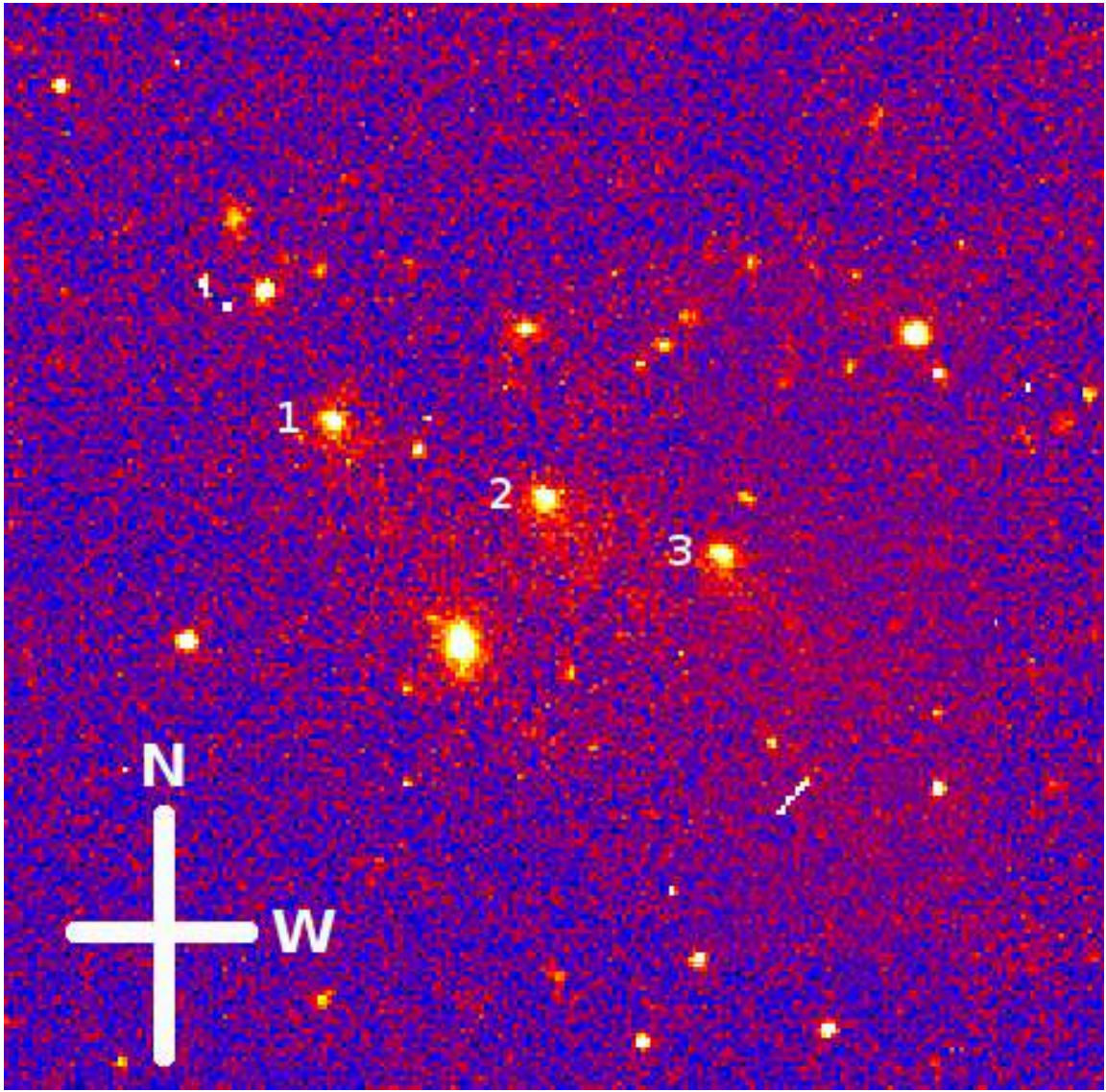}
\caption[Unknown Comet]{
	\label{fig1}
	\label{lastfig}			
	A combination of three images, taken 45 minutes apart, of the unknown comet observed 2007 September 14. The comet moves East to West in the sky. A tail of $\sim10^{\prime\prime}$ is seen to the right (West) of the object. See online supplementary material for an animation of this object.
	}
\end{center}
\end{figure}

\begin{figure}[t!]
\begin{center}
\includegraphics[width=6in]{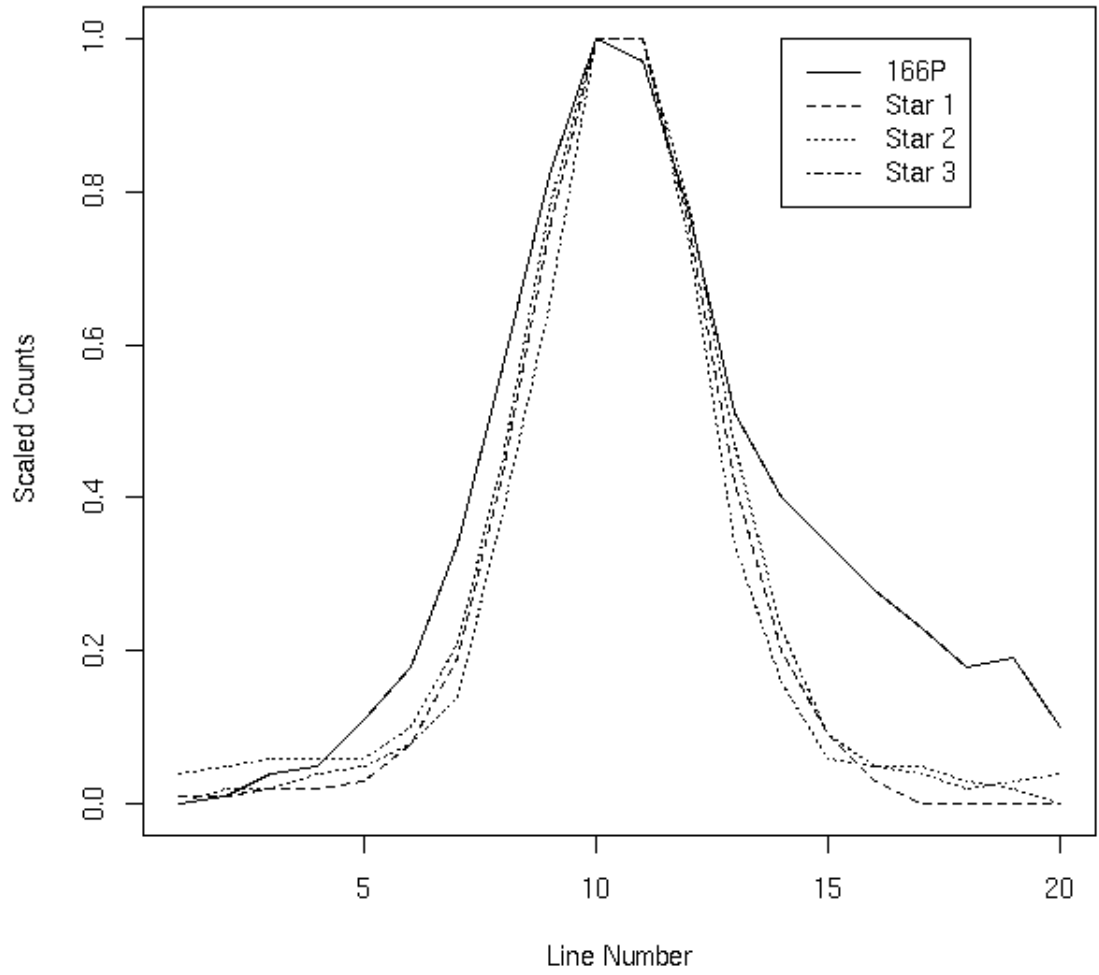}
\caption[Unknown Comet]{
	\label{fig1}
	\label{lastfig}			
	Brightness profile of 166P (solid line) and three comparison stars (dotted lines).
	}
\end{center}
\end{figure}

\begin{figure}[t!]
\begin{center}
\includegraphics[width=6in]{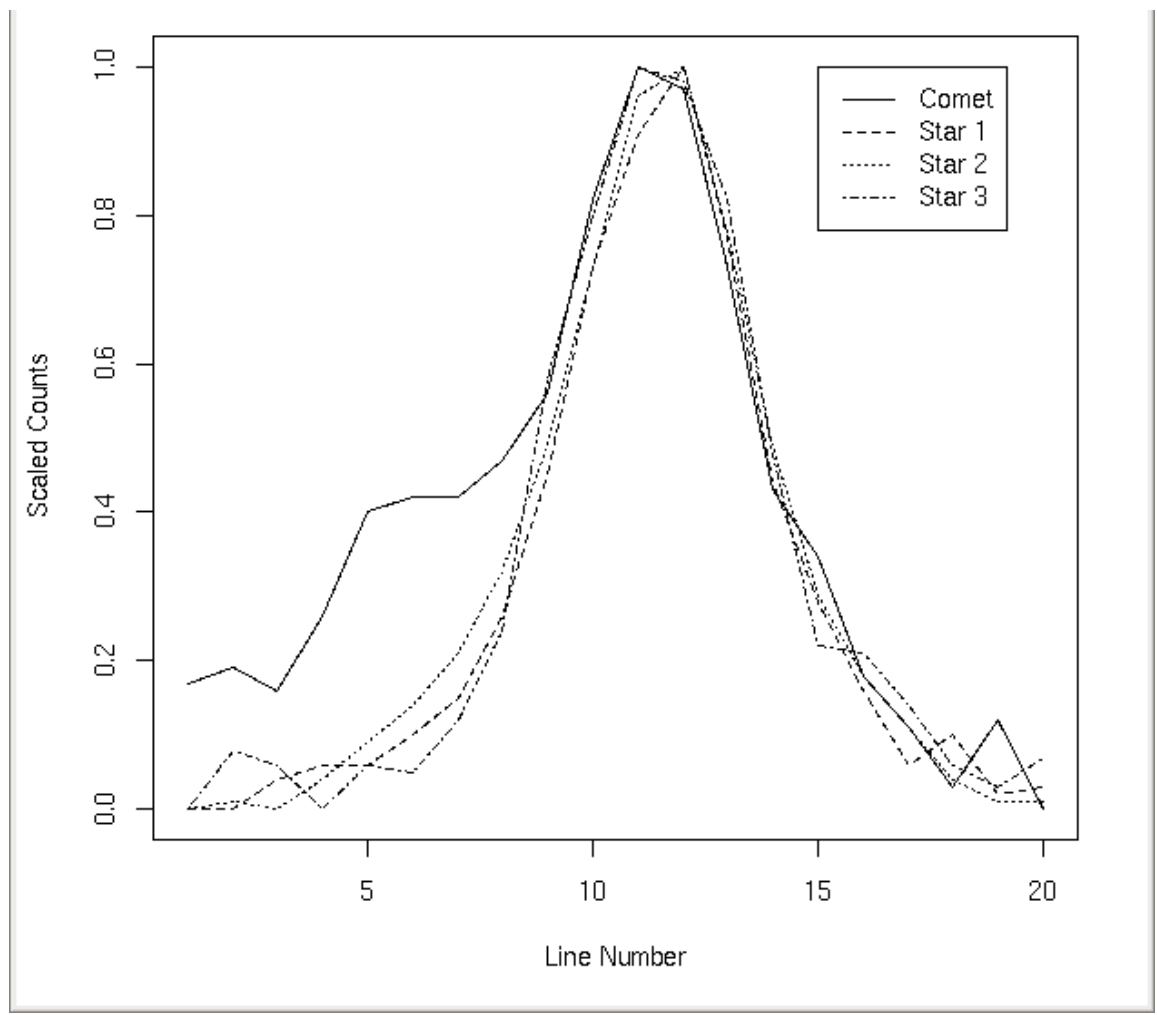}
\caption[Unknown Comet]{
	\label{fig1}
	\label{lastfig}			
	Brightness profile of the unknown comet (solid line) and three comparison stars (dotted lines).
	}
\end{center}
\end{figure}

\begin{figure}[t!]
\begin{center}
\includegraphics[width=6in]{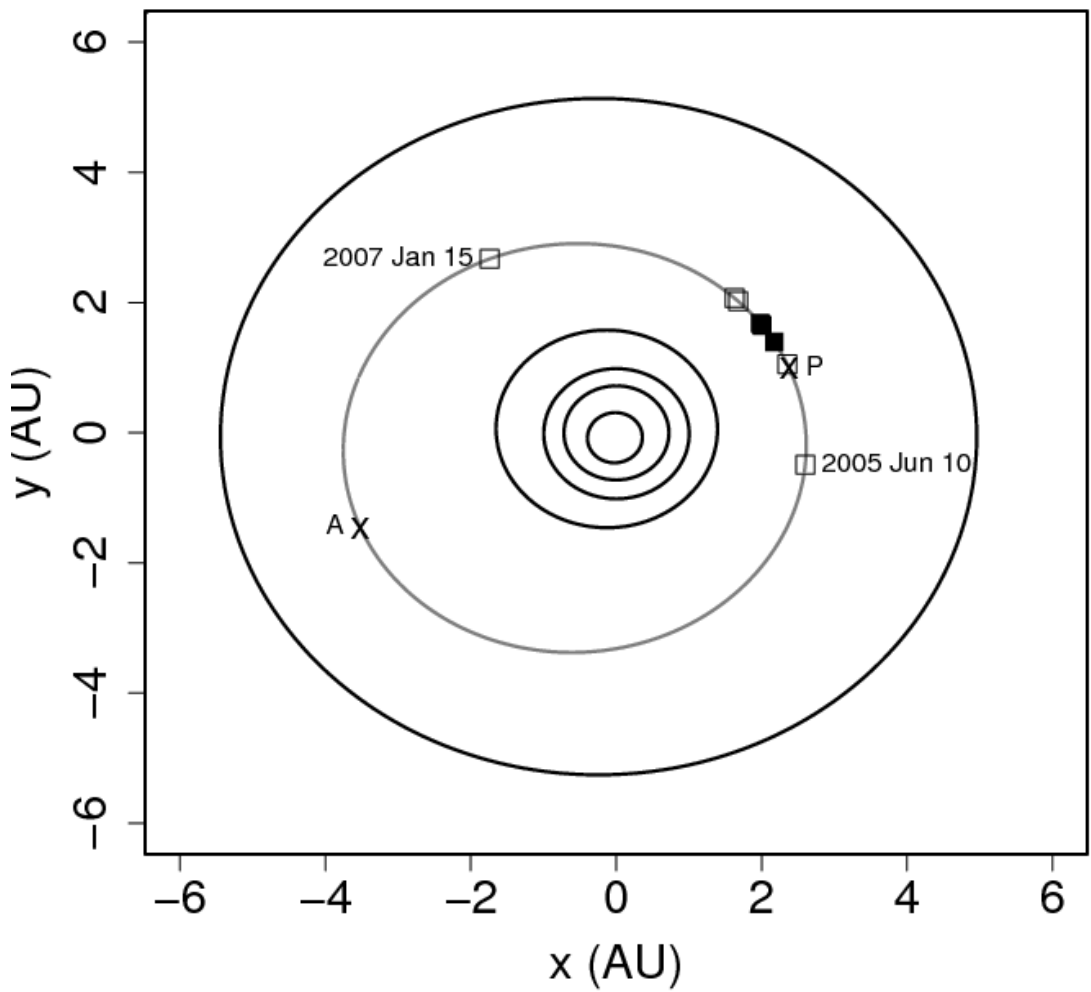}
\caption[Observations of 176P]{
	\label{fig1}
	\label{lastfig}			
	Orbital diagram of 176P showing its inactive and active phases. Orbits of Mercury, Venus, Earth, Mars, and Jupiter are shown in black, while the orbit of 176P is in gray. Aphelion (A) and perihelion (P) positions are also marked. Solid and open squares show positions where 176P is active and inactive, respectively. The two positions marked with dates are from the current study; all other positions are from \citet{HJ2006} or \citet{H2007}.
	}
\end{center}
\end{figure}

\end{document}